\documentclass[a4paper,12pt,reqno]{amsart}
\pdfoutput=1

\usepackage{a4}
\usepackage{amsthm}
\usepackage[T1]{fontenc}
\usepackage[utf8]{inputenc}
\usepackage[british]{babel}
\usepackage{todonotes}
\usepackage[unicode,psdextra]{hyperref}
\usepackage{tikz}
\usetikzlibrary{decorations.pathmorphing,shapes}
\usetikzlibrary{arrows.meta,calc}
\def\centerarc[#1](#2)(#3:#4:#5)
    { \draw[#1] ($(#2)+({#5*cos(#3)},{#5*sin(#3)})$) arc (#3:#4:#5); }

\makeatletter
\@addtoreset{definition}{section}
\@addtoreset{equation}{section}
\makeatother

\makeatletter
\let\@wraptoccontribs\wraptoccontribs
\makeatother

\allowdisplaybreaks[4]

\begin{document}

\title[The Reasonable Ineffectiveness of Aesthetics in Particle Physics]{The Reasonable Ineffectiveness of Aesthetics in Particle Physics}

\author[J. Branahl]{Johannes Branahl\textsuperscript{1}}
 
\address{\textsuperscript{1}Philosophisches Seminar der
Universit\"at M\"unster \hfill \newline
Domplatz  23, 48143 M\"unster, Germany \hfill \newline
{\itshape e-mail:} \normalfont
\texttt{j\_bran33@uni-muenster.de}}

\begin{abstract}
In recent years, criticism of the methodology of particle physics beyond the Standard Model has increased, diagnosing too much reliance on aesthetic criteria for theory development and evaluation. Faced with several decades of experimental confirmation of all theories lacking, we subject four aesthetic criteria - simplicity, symmetry, elegance, and inevitability -, regularly mentioned in theory evaluation, to critical examination. We find that these criteria, all of which can be reduced to a desire for simplicity, have repeatedly misled modern particle physics. This is largely due to the lack of metatheoretical permanence of a uniform conception of simplicity. The reductionist claim of particle physics - the search for simple fundamental principles in a complex world - will be worked out as the reason why this discipline is particularly susceptible to the aesthetic appeal of simplicity. Thus, compared to disciplines dealing with complex phenomena, aesthetic criteria are much more frequently applied, exposing particle physics to the risk of missteps and dead ends.
\end{abstract}


\maketitle
\markboth{\hfill\textsc\shortauthors}{\textsc{{The Reasonable Ineffectiveness of Aesthetics in Particle Physics}\hfill}}

\section{Introduction}
Several decades ago, physicist Eugene Wigner penned an article on the "Unreasonable Effectiveness of Mathematics,"  which one encounters in the natural sciences. Astonishingly, according to him, mathematics has not only been predictive but has also paved the way to new theoretical achievements as a reliable guide (Wigner 1960). Throughout many epochs, aesthetics in theory often joined mathematics as a guiding light, yet it regularly fell short. Aesthetic criteria for formulating and evaluating theories can be seen as metatheoretical criteria, triggering an aesthetic sense when fulfilled. Their validation is derived from the old, but unprovable assumption that beauty and truth go hand in hand in natural sciences (\textit{pulchritudo splendor veritatis}). One cannot definitively explain the unparalleled success of mathematics across all epochs, it remains "unreasonable." However, there are clear reasons for the failures of aesthetics. Thus, the aim of this article is to elucidate the "reasonable ineffectiveness" of aesthetics, particularly in a field marked by many failures: particle physics beyond the Standard Model (BSM physics for short).

The frequent appearance of the subjective concept of aesthetics in physics, which prides itself on mathematical precision and objective description of the world across all scales, initially seems perplexing. The aesthetic content of a set of equations revealing the fundamental principles of the Universe appears accessible only to a small group of initiates, unlike in arts, poetry or music. Moreover, the inclination toward aesthetic vocabulary is particularly noticeable in fundamental physics, while in other disciplines, only a few marginal notes on beauty can be found (such as the double helix structure of DNA, see Watson 1968). For some, the aesthetics of physical theories indicate the presence of a \textit{creator spiritus} or a mitigated creative principle pervading the Universe (Boltzmann 1893), while for others, aesthetics serve as a criterion to pursue or discard theories: Withrow (1967) analyzed Einstein's aversion with respect to ugly equations, convinced that beauty was an indicator for important results in theoretical physics."

Besides beauty in general, concepts such as simplicity, symmetry, elegance, inevitability, and naturalness are often found in the aesthetic characterization of particle physics achievements, especially over the last hundred years. Not only does beauty as a whole resist a unified definition in the history of philosophy (see Carroll 1999), but these mentioned five concepts also largely elude a uniform definition or objectification when applied to physics. In each quotation and theory, they may be interpreted and understood differently. This article cannot provide uniform definitions either. Sometimes, the criteria seem to be flowery descriptions of clear methodological rules of theoretical physics, while at other times, they appear contradictory or inconsequential. In this article, we aim to analyze the rôle of the first four criteria in the research programme of contemporary BSM physics and its negative influence by aesthetic evaluation. While the fifth criterion, naturalness, has already been extensively discussed in the philosophy of physics\footnote{Contributions to the naturalness debate include (Giudice 2008, Wells 2013 Dine 2015, Hossenfelder 2017, Williams 2015, 2019, Wallace 2019, Bain 2019, Franklin 2020, Koren 2020, Fischer 2023, Branahl 2024).}, as it presumably has most shaped the research programme to date, there have been only initial impulses in the search for other indicators of beauty in particle physics. For instance, Hossenfelder's criticism (2018) became already accessible to a broader public. Given the ongoing stagnation beyond the Standard Model, a deepened discussion of the rôle of aesthetic criteria is a crucial component for crisis diagnosis\footnote{Some ideas detailed out in this article already appear in Hossenfelder's (2018) popular science book, mostly with respect to simplicity, but lack a concrete distinction of definitions given the scope of her work.}.

In Chapter 2, we will examine the four criteria of simplicity, symmetry, elegance, and inevitability in detail. Each of the four subchapters will attempt to synthesize the central concepts and definitions of the respective criterion. Subsequently, the criteria will be contextualized within particle physics, where we will analyze the limits and failures of the four criteria. We briefly look at the connection of simplicity to naturalness in Chapter 2.5. Chapter 3 will condense the insights from the previous chapter into two theses. First, the four criteria do not exist in isolation but are strongly intertwined and ultimately reducible to simplicity. The desire for simplicity, stemming directly from the reductionist claim of fundamental physics, makes this discipline particularly receptive to the inclusion of aesthetics in the search for scientific progress, thereby exposing it to the danger of errors and dead ends more than any other discipline: dangers arising from the unjustified characterization of aesthetic criteria as guiding principles. Second, this characterization is unjustified because the criteria lack a property that we want to call \textit{metatheoretical permanence.} While the effectiveness of mathematics relies on the metatheoretical permanence of the mathematization of the explananda, the justifiable ineffectiveness of aesthetics is partly due to the changing aesthetic ideals, as described by McAllister (1996) and brought into the context of BSM-physics by Hossenfelder (2018). Chapter 4 summarizes the findings of the article.

\section{Four Central Aesthetic Criteria in Particle Physics}
 
\subsection{Simplicity}
 In the history of science, various concepts of simplicity have been used as criteria for theory selection and (pre-experimental) evaluation. Firstly, we must argue for the not immediately evident connection between simplicity and aesthetics to be able to attribute the failure of some BSM theories to an excessive reliance on aesthetics as a guide for (approximately) true theories—or at least to declare this unwarranted trust as a potential contributor to the current crisis. To define the demand for simplicity in BSM physics, we must make some conceptual distinctions afterwards.

Let's first consider the notion of simplicity according to Dirac, who dogmatically prescribes preferring a beautiful theory over a simple one in case of doubt: According to him, a theorist should always strive for mathematical beauty when formulating the laws of nature. While the conditions for beauty and simplicity are often the same, beauty should be favored otherwise (Dirac 1939). Hence, from some viewpoints, simplicity is not necessarily connected to aesthetics. Nevertheless, aesthetics allows for a plausible ascription of intrinsic value to simplicity, which can justify the demand for simplicity (Derkse 1992). To avoid theological or metaphysical justifications with their far-reaching consequences, Sober suggests in addition: "Just as the question ‘why be rational?’ may have no non-circular answer, the same may be true of the question ‘why should simplicity be considered in evaluating the plausibility of hypotheses?’" (2001, p.\,19). This merely requires an intrinsic value of simplicity, for which aesthetics is a prominent candidate. But where does the aesthetic appeal of simple theories come from? To answer this, let's introduce some definitions.

Firstly, we'll delineate the notions of simplicity relevant to the problem from what we want to term as \textit{naive} simplicity. Haldane (1927) early noted the subjectivity inherent in ascribing simplicity. Theory evaluation through the aesthetic criterion of simplicity is nothing more than subjective criticism akin to that of artworks or poetry. He compares the simplicity of the wave equation with the simple verbal description "it oozes" in the eyes of lawyers and physicists. Their perception varies from discipline to discipline. However, we'll focus on the intradisciplinary attribution of simplicity—few non-physicists would presumably label the equations of the Standard Model as simple. McAllister (1996, p.\,107) attempts to define the simplicity of equations: In a polynomial dependence, small powers shold be preferred over large ones, generally whole numbers over real ones. In the case of BSM physics and its highly developed mathematics, which involves entirely different structures, this criterion must be relinquished—thus, it can be often argued from an external perspective using a conception of \textit{naive} simplicity.

Next, we must differentiate between \textit{contextual} and \textit{idealized simplicity}. Contextual simplicity refers to preferring the simpler theory among theories that can explain the same phenomena or observations. This means that if two or more theories provide equivalent explanations for a phenomenon, we prefer the one with fewer assumptions or complexity. William's of Ockham scholastic maxim- a law of parsimony already encapsulates this in the Middle Ages: The scientist should not make life harder than necessary and should adhere to the principle of ontological parsimony. Also, the more modern principle of Mach (1886, p.\,586) of scientific simplicity as a minimization problem (of the smallest mental effort while considering a complete set of facts) proved to be a helpful guiding principle. One cannot conclusively justify why this principle of parsimony repeatedly prevailed in the history of science. However, we will find that today's BSM physics cannot clearly apply this principle, as either competing notions of simplicity—and resulting competing theories—exist, or no contextual simplicity is demanded, but a more difficult to justify property: Under \textit{idealized} simplicity, we understand the transfigured desire for an even simpler theory, even if there are no direct competitors. In this case, one wishes for ever greater simplicity at any cost, even if existing, indispensable theories exhibit empirical adequacy.

In the context of today's BSM physics, the restriction of contextual simplicity, the principle of "not making it harder than necessary," deserves special emphasis. Consider, for example, Hawking's (2002) aesthetic judgment of the Standard Model, which is based on idealized simplicity: "First of all, it is ugly and ad hoc. The particles are grouped in an apparently arbitrary way, and the Standard Model depends on 24 numbers, whose values cannot be deduced from first principles (...). Can it be Nature's last word?" This quote reveals the hope that ultimately, a minimal set of input parameters exists from which everything else can be derived based on universal principles. In other words, the Standard Model is perceived as ugly because 24 free parameters are not fundamental enough—or in other words, not simple enough. However, there has been no simpler alternative for over half a century, and predictions of the Standard Model are regularly confirmed with the highest precision.

Of course, there is no doubt about the provisional nature of the Standard Model in the absence of quantum gravity and for many other reasons. However, the history of science presents a completely different picture regarding simplicity than Hawking imagines: From the doctrine of the Four Elements—the beginning of a series of efforts of cosmic inventory—the Standard Model of elementary particles with its non-abelian gauge theories of relativistic quantum mechanics and the Higgs mechanism, as well as General Relativity, emerged as pillars of our current worldview. Even the relatively simple notion of a point-like proton with the quantum numbers of charge, mass, and spin from the first half of the last century has been replaced by a substructure of three valence quarks and a sea of constantly emerging and disappearing quark pairs and gluons, whose permanent interaction appears arbitrarily complicated, at least in the perturbative picture. A simple classification has been replaced, through its refinement, by continuously more quantum numbers, objects, and parameters, but above all by an increasingly sophisticated mathematics. Similarly, Newton's theory of attracting masses, whose force decreases with the square of the distance, has become a system of 16 coupled, nonlinear differential equations, involving four-dimensional differential geometry. Regarding simplifications in Hawking's sense, one can justifiably extrapolate: It will probably never be as simple again as it is today. This reflects the whole problem of the \textit{idealized} variant of simplicity. The ultimate worldview, if it exists, could be much less beautiful than Hawking already describes the Standard Model due to the lack of such simplicity. Hossenfelder considers the possibility of an ugly Universe already in the title of her book (in the German translation, 2018). However, the development of the Standard Model has not violated the demands of William of Ockham and Mach, which only represent the call for contextual simplicity. However, the search for idealized simplicity beyond the Standard Model may turn out to be fundamentally unsuccessful.

Let us now turn to \textit{contextual simplicity}, which also brings problems in BSM physics. In order to apply the principle of ontological parsimony, the ontology underlying the theory must be clarified. A frequently advocated division of simplicity distinguishes between syntactic (number and complexity of hypotheses) and ontological simplicity (number and complexity of postulated entities)—two ideas of simplicity that usually cannot be increased simultaneously. An example often used is the postulate of Neptune to explain the perturbations of the orbits of already discovered planets: It increases the number of entities in the solar system and thus makes it more complex, but it maintains the (simple) theory of celestial mechanics. According to Quine (1966), a balance is struck between the simplicity of the world (ontology) and the simplicity of the theory (ideology) addressing the same idea of syntactic and ontological simplicity. While this distinction may have been appropriate in the history of science, in modern particle physics, the boundaries between these forms of simplicity blur. Instead, we propose a distinction between simplicity with respect to an \textit{ontology of information} (equivalent to ontological simplicity, adding information about new particles in the form of a set quantum numbers) and  simplicity with respect to an \textit{ontology of relation}\footnote{In the sense of an ontological structural realism as proposed by Ladyman 1998, Chakravartty 2012). BSM physics deals with identifying the mathematical structures of our world, which can be considered ontologically fundamental, unlike the preliminary and approximate mathematical structures of celestial mechanics. Of course, also a set of quantum numbers might be treated as a mathematical structure, ending the illusion of materialism.} (how entities are related and interact, e.g., via internal symmetry groups), each of which minimizes simplicity.

This distinction has the following background: The postulate of Neptune did not change the laws of celestial mechanics because it cannot be understood as an ontologically fundamental entity, but is an astronomical contingent. (B)SM physics, on the other hand, deals with elementary particles that are considered irreducible at least according to the current state of theory building. The postulate of new particles (extension of the ontology of information) implies a more complex theory, too, that embeds the new particle into the mathematical structures of quantum field theory e.g. through an additional gauge symmetry (extension of relational ontology through additional \textit{first principles}). Conversely, an extension of the theory through additional (external and internal) symmetries automatically implies the existence of new particles. Such an entanglement is no longer representable by the classical distinction between syntactic and ontological simplicity.

 In the history of the Standard Model, it was possible to simultaneously minimize both ontologies while producing an empirically adequate theory: For instance, Gell-Mann (1964) was able to reduce the particle zoo of hundreds of mesons and baryons—initially assumed to be elementary—to their fundamental constituents of three quarks through the postulate of a $SU(3)$ symmetry. The aesthetics of simplicity in both respects may have strengthened confidence in the quark model before its experimental confirmation\footnote{String theory initially seems to fulfill both forms of simplicity—a single entity (strings) with a single principle (the excitation of strings). However, it is a theory whose ontological parsimony in the microcosm of strings as the only fundamental component pays the high price of ontological generosity in the macrocosm: the price of 11 or 26 dimensions and approximately $10^{500}$ parallel universes (cf.\,Ibáñez, Uranga 2012).}.

However, in BSM physics, one faces the following challenge: Some theories postulate an additional first principle in the form of a symmetry to explain certain phenomena and accept an explosive increase in new particles. We will consider Supersymmetry (SUSY) and proposals for a Grand Unified Theory (GUT) of the three interactions understood on a quantum level (without gravity) below. In contrast, other theories (like Minimal Models) seek to prevent this by minimizing the number of new particles—at the expense of less aesthetic first principles. Empirical evidence supporting any of these theories is completely lacking so far. Lacking other evaluation criteria is a central reason why the aesthetic criterion of simplicity is used for theory selection.

The theory of SUSY (see for an introduction: Martin 1996) postulates a single external symmetry that gives rise to the existence of so-called superpartners of particles. The superpartners have half-integer spin (fermionic particle partners) when the Standard Model particles have integer spin (bosonic particles), and vice versa. GUTs combine the three internal symmetries of the Standard Model, $SU(3) \times SU(2) \times U(1)$ into a single symmetry group (among others, $SU(5)$, $SO(10)$, or $E(8)$). Both ideas initially seem aesthetically appealing due to their simplicity in a relational ontology. However, is the much-vaunted beauty of SUSY justifiable with its up to 105 free parameters? Is it ontologically parsimonious to double the particle content of the Universe? Are a total of 45 gauge bosons of a GUT, generated by the $SO(10)$ Lie algebra (Pati, Salam 1975), simple? Or the 248 dimensions of the exceptional Lie group $E_8$, as proposed in Lisi's (2007) GUT? Since Minimal Models have also not received empirical support, no statement can initially be made about whether the loss of simplicity in the ontology of information could be problematic and speaks against its empirical adequacy.

We have seen that for BSM physics, neither the simplicity of mathematics (naive simplicity) nor the economy in introducing new objects and parameters (idealized simplicity, simplicity of an ontology of information as demanded by Hawking) can be an aesthetic guiding principle. These helpful guidelines (over an extended period) seem to have been exhausted due to the unexpectedly complex nature of the microcosm. Even the relatively late-found Klein-Gordon (bosons) and Dirac equations (fermions) of relativistic quantum mechanics enjoyed high aesthetic popularity (cf.\,Kragh 1990) due to their mathematically extremely simple form. This simplicity is surprising in light of the generally challenging and extensive mathematics of this discipline. Today, these epochs are over. Instead, only in the definition of simplicity through the reduction of the world to a few first principles (relational ontology) its semantic adequacy and aesthetic content is restored in contemporary BSM physics (but might change again within years!). Thus, BSM theories may bring as many extensions in terms of an ontology of information as well as mathematical complications as necessary, as long as the demand for contextual simplicity is preserved.

However, the problem of theory evaluation remains unresolved—van Fraassen (1980) calls systematization through first principles a pragmatic reason for accepting theories but does not see it as a reason to consider them true. Morrison (2000, p.\,138) writes supportively about unifying theories: "It is quite a different matter to say, as Friedman and others do, that the unifying structure should be interpreted realistically simply in virtue of that function, or to claim that inference to the 'unified explanation' is a legitimate methodological practice." McAllister adds to these thoughts: "So appeal to the notion of unifying power establishes neither that the simplicity of theories is a purely empirical property of them nor that the notion of simplicity has no aesthetic aspects." (1996, p.\,111). In a similar pattern, Salimkhani (2021) argues regarding GUTs in particular: In this context, he distinguishes between physical argumentation (there: internal explanation) and metatheoretical assumptions (there: external explanation) and provides reasons why the unification of gauge forces can be considered a by-product of physical research. Building on Maudlin (1996), he mentions the determination of the Weinberg angle and the explanation of the quantization of the charges of all elementary particles as multiples of the electron charge as physical reasons to strive for unification. We can principally agree with this\footnote{In addition, there is the natural consequence of a seesaw mechanism to explain neutrino masses (Senjanovic 2014).}, even if these phenomena do not necessarily require an explanation by a GUT but could be explained by future (simpler?) candidates. In any case, the crucial insight is the following: Only the explanation of physical phenomena can ultimately be used as a valid argument for adhering to theories\footnote{In contrast to Hossenfelder, we therefore see the search for a GUT as a benign physical problem, as long as candidates reveal clues for experimental verification.}.
 
A way out, where the simplicity of theory could go beyond a "pragmatic reason for acceptance," would be a historically clear and unchanging notion of simplicity. However, this seems hopeless: "Simplicity is largely a matter of historical background, of previous conditioning, of antecedents, of customary procedures, and it is very much a function of what is explained by it," von Neumann wrote (1955).

This general characterization of simplicity is easy to support. A small number of universal principles seem to correspond more closely to the concept of simplicity in light of the current state of knowledge in particle physics than the reduction of the world to a small number of causal interactions (here due to fewer fundamental entities). As McAllister (1996) points out, both notions are different approaches to the unification of phenomena with different aesthetic properties, and no one approach can be attributed absolute superiority in accessing the world. Indeed, in the history of science, each of these approaches has temporarily seemed promising\footnote{Even before the advent of natural science, there were competing notions of order in the world through simplicity: the postulate of indivisible particles, all having the same properties (atomism according to Leucippus and Democritus - from today's perspective, an ontology of information) or through symmetries as first principles (symmetries of Platonic solids - a relational ontology)}. While Newton's worldview, with the unification of celestial and terrestrial mechanics, made a significant contribution to the simplicity of the world through the omnipresence of the causal law of gravitation, Leibniz's efforts towards simplification through first principles such as continuity, conservation, and relativity were supposed to succeed (cf.\,Boullart 1983). 

The preceding discussion illustrated the failure of simplification through low complexity in an ontology of information in particle physics. This initially successful course of unifying phenomena, initiated by Newton, reaches its limits in the extensive catalog of particles and parameters in the microworld, and is even reversed. Unification in the microworld, as it stands today, means a richness of phenomena, but a reduction of principles. Several centuries of Newtonianism now seem to be replaced not only from a physical but also, with a time delay, from an aesthetic perspective: in the latter case, by the change in the aesthetic ideal of simplicity due to the progress in standard model physics, or more precisely: in its symmetries.

\subsection{Symmetry}

While simplicity in physics has resisted a uniform definition, and we initially defined it in BSM physics through the reducibility of the phenomena of this world to a few fundamental principles, symmetry appears to be a considerably clearer aesthetic criterion. The objective content of this aesthetic criterion lies in the mathematically precise definition of symmetry in physics: a transformation that leaves the physical system unchanged (invariance). On one hand, the presence of symmetries is an immediate simplification of the system since it restricts the information needed to describe it. In the context of particle physics, but also physics as a whole, certain symmetries reveal themselves as fundamental principles, according to which simplicity, in the sense described earlier, arises. While one does not necessarily have to find symmetries aesthetically appealing to make use of them in theory-building, it is reasonable to assume that in several chapters of the history of physics, the aesthetic allure of symmetries has motivated researchers to pursue their theories more fervently and not lose sight of them. Regarding Maxwell's equations of electromagnetism, one reads: "It would seem that the symmetry of these equations and the aesthetic appeal that this symmetry generated must have played an important role for Maxwell in his completion of these equations" (Penrose 1974, p.\,271).

The allure of symmetries is understandable: we encounter them in everyday life when judging the beauty of faces (Rhodes 1998) and in architecture since ancient Greece. Indeed, the desire for symmetry dominated the Greek philosophy to such an extent that it became a Western ideal of beauty (cf.\,Pollitt 1974). However, it is noteworthy that in addition to the subjective positive perception when considering symmetries, there is a factual relevance of symmetries anchored in the basic principles of the cosmos. Perhaps we only know this latter fact because of the former—an evolutionary accident that occasionally allows beauty to serve as a guide to truth: it would be a welcome coincidence between mathematical structures of the Universe and human aesthetic perception, from which, of course, one cannot infer a beauty of the Universe independent of the observer, but certainly some simplicity. With symmetries, we find a mathematical refinement and conceptual narrowing of the criterion of simplicity, which comes into question as a guiding principle in theory-building. Not least, the gauge symmetries of the Standard Model, the chiral symmetries of fermions, the octets and decuplets of baryons, and the symmetric Higgs potential in the unification of electromagnetic and weak interactions support this statement.

However, to successfully apply symmetries, it must be clear whether they are fundamental. A desire for perfectly circular planetary orbits due to their higher symmetry compared to ellipses may have been a prevailing aesthetic ideal in cosmology during the Renaissance. However, today we understand ellipses as a by-product of planetary formation, which has no relevance for fundamental physics. Furthermore, it is crucial to acknowledge that symmetry can at most be a necessary condition for a significant theoretical addition in BSM physics, but not a sufficient one. This is already evident in the competition for the correct symmetry of the GUT or the absence of any confirmation for supersymmetric particles. Finally, it should be noted that once again their rôle at the most fundamental level of the cosmos has been quietly questioned for some time, which challenges the (global) symmetric nature of the fundamental principles by which we understand simplicity in BSM physics (Kallosh et al., 1995; Banks and Seiberg, 2011). Perhaps one day we will see the symmetries of the Standard Model as similarly insignificant as the symmetries of the solar system.

In conclusion, we note that even the restriction of simplicity to mathematically clearly defined principles (symmetries) does not guarantee a guiding light toward truth. However, it is clear that the outstanding importance of symmetries can justify a certain inclination towards a more symmetric candidate in the phase of theory-building, as long as it is not the only justification used. The abundance of symmetries in many BSM theories, few if any of which accurately reflect reality, should be a sufficient warning. Still, this does not mean that the search for symmetries should not be anchored in the modus operandi of this research programme. However, it appears as a weaker criterion for truth if the symmetry of the theory is not discovered accidentally as a deeper principle during development (recall Gell-Mann's Eightfold Way, which hinted at the existence of quarks), but rather declared as a construction principle and symmetries are proposed ad-hoc. In the following, we will see that aesthetic feelings arise, for various reasons, even more strongly in retrospect on already developed and deemed correct theories. From the rather objective aesthetic criterion of symmetry, it often follows the ex-post impression of \textit{inevitability}, while from the general simplicity of a theory, despite its phenomenological richness, \textit{elegance} is regularly spoken of. Its relevance for an objective theory evaluation will be examined in the next section.

\subsection{Elegance}
 
The concept of elegance is often appropriated by mathematicians and physicists to express their retrospective aesthetic delight over a theory and to highlight its beauty as a reason for its correctness. (Taylor 1966), for instance, wrote about Einstein that the elegant beauty of general relativity sufficed to believe in its truth. Hossenfelder (2018, p.\,127) describes elegance as a combination of the simplicity of the theory and the surprising richness and complexity of the phenomena it predicts—a surprise inherent in the eureca-effect. The subjectivity of this criterion is hardly questionable. Hossenfelder also argues against attempts to objectify it by Dawid (2013), who tries to formalize elegance as the "unexpected explanatory conclusion" and establish it as a valid post-empirical criterion but fails due to the subjective experiential content of the unexpected (ibid.).

Elegance thus goes beyond the mere explanatory power of a theory; it addresses the aesthetic appeal of the unexpected. Hossenfelder's extension stands in the tradition of Zahar, who, as a student of Lakatos, defined a class of new facts that were actually already known but \textit{unexpectedly} emerged as a consequence of a new, simpler theory. He thus complemented the essay of his teacher, which discussed why the Copernican model seemed more persuasive than the Ptolemaic one (Lakatos 1978, p.\,168). Related concepts, such as the aforementioned syntactic simplicity or consilience according to Thagard (1978), leave aside the personal expectations of researchers and thus possess a higher degree of objectivity in terms of theory evaluation. The close intertwining of simplicity and elegance is explained by von Neumann: "(...) the criterion of success for such a theory is simply whether it can, by a simple and elegant classifying and correlating scheme, cover very many phenomena, which without this scheme would seem complicated and heterogeneous, and whether the scheme even covers phenomena which were not considered or even not known at the time when the scheme was evolved." (1947, p.\,6).

In the aforementioned work by Taylor (1966), the elegance of general relativity is praised. Indeed, from the parsimony of a few basic assumptions about the world (principle of equivalence, Mach's principle, principle of relativity) an immense wealth of new insights arises: the various solutions to Einstein's field equations explain the precession of Mercury's perihelion, describe gravitational lenses and the general deflection of light by stars, predict gravitational waves, suggest the possibility of a Big Bang and the dynamics of the expanding Universe, describe the existence of black holes, bring forth wormholes as unphysical solutions, and increase the precision of GPS-based navigation. This list could be extended, and many conclusions drawn from Einstein's field equations were unexpected, in the case of the expansion of the Universe even for Einstein himself (cf.\,Gamow 1970, p.\,44). Elegance apparently hints at the possibility that a very fundamental equation must exist for a subfield of physics, from which various more complex phenomena of the same subfield can be derived. The accompanying aesthetic sensation, however, seems ultimately to be a retrospective amazement at the confirmation of reductionism, which the surprised physicist must originally have assumed to successfully reduce the diversity of the world to a few equations: a circular argument. There is nothing inherently wrong with this understanding of elegance, as long as elegance is not overinterpreted as a guide to truth.

As from the parsimony of assuming a fundamental symmetry between bosons and fermions (SUSY) also arose a considerable richness in particle physics, we return to the problems of BSM physics. The symmetric extension of QFT was maximized by supersymmetry according to (Haag, Łopuszański, Sohnius 1975)—its demand was not ad-hoc. The following unexpected conclusions could be drawn during the elaboration of SUSY: SUSY appears as a natural byproduct in string theory (Ramond 1971), provides (each with certain parameter choices) a candidate for dark matter, lets the coupling constants of the three gauge forces converge to a point, and resolves the large, thus unnaturally perceived perturbative corrections to the Higgs mass via scalar loop corrections (assuming that these problems are considered solvable by particle physics). A prime example of the definition of elegance at the beginning of the section. Yet, no SUSY particles have been discovered for most suitable regions in the parameter space.

The aesthetic criterion of elegance thus carries the risk of proving to be a false friend—an unreliable guide in the search for truth. A good and confirmed theory will often retrospectively be considered elegant, but an elegant theory does not necessarily need to be confirmed. In Einstein's case, these conclusions were experimentally confirmed. In the case of SUSY, clinging to its elegance may have contributed to an uncritical dogmatism in particle physics. The necessary but missing findings of particles on the GeV to TeV scale have led to the fact that the mentioned explanations by SUSY are no longer readily possible, but not to stop working on SUSY altogether. A Bayesian perspective may help explain the failure of the elegance criterion. SUSY is a simple (due to only one symmetry requirement) but vague theory (due to its high-dimensional indeterminate parameter space). The options for parameter configurations are so numerous that initially a high number of potentially true-positive facts of the theory emerges. This is interpreted as theoretical richness arising from simplicity, and it initially seems unlikely that SUSY is false. However, the Bayesian concept of conditional probabilities presents a much higher number of false-positive results (which increase in number with each experiment of the ATLAS Collaboration) against the potentially true-positive results of SUSY. This holistic, Bayesian view shows, on the one hand, that the surprise about the potentially true-positive results is unjustified and deceptive. On the other hand, it also refines its theory of science classification, which in the classical conception would consider SUSY as not yet falsified, as there are still unreached regions of the parameter space. However, the concept of conditional probabilities considers SUSY as "probably falsified" and relativizes the aesthetic allure of elegance.

\subsection{Inevitability}
The concept of inevitability arises when the retrospective significance of certain symmetries is recognized as fundamental principles that inevitably reveal the specific form of the laws of nature. This recurring success story of symmetries in physics is evident, for example, in the conservation laws of classical physics, which mathematically follow from the invariance of physical systems under symmetry transformations (Noether 1918). Similarly, gravity in General Relativity follows from the symmetry under the transformation of reference frames. In its quantized form, Einstein's equivalence principle arises from the symmetry property of the graviton (spin-2 boson). And finally, the fundamental interactions at the quantum level arise from the gauge symmetries of the Standard Model: the $SU(3)$ symmetry under rotations in color space makes the concrete manifestation of the strong nuclear force in the world inevitable. Similarly, arguments are made for the symmetries $SU(2)$ in isospin space. The non-abelian $U(1)$ symmetry of QED also rules out photon-photon interactions. The equation for such a 4-particle interaction has no solutions. Ultimately, this is why we can see over large distances; the light rays do not interact with each other on their way to us.

Let's read what Weinberg (1993, p.\,107) writes about this inevitability: "In listening to a piece of music or hearing a sonnet one sometimes feels an intense aesthetic pleasure at the sense that nothing in the work could be changed, that there is not one note or one word that you would change. (...) The same sense of inevitability can be found (...) in our modern Standard Model of the strong and electroweak forces that act on elementary particles." This analogy does not do justice to the objective nature of this principle. A Beethoven symphony may give the feeling that small changes to the score could diminish the value of the work. This subjective feeling of inevitability in the works of the most significant composers and physicists may be a sign that they mastered their respective discipline to perfection—having a deep intuition about what the perfect composition or theory must look like. A change in the theme of a sonata or in the arrangement of figures in the School of Athens is conceivable; it can remain a completed sonata or a remarkable piece of artwork. However, changes in the equations of the Standard Model do not allow this. While the \textit{feeling} of inevitability expresses an idealized, transfigured perspective on great artists, the \textit{principle} of inevitability guarantees the internal consistency of physics—an expression that we can rationally understand the world through science. The subsequent proof of the inevitable consequence of our laws of Nature from certain principles is nothing but the idea of a comprehensible world. Equating aesthetic sensibility in theory construction and evaluation with inadequate subjective criteria is thus a false judgment in the case of the inevitability of the theory (ex post) and occasionally in the symmetry of the theory (ex ante). Nevertheless, aesthetic sensibility is a by-product in the construction of good physical theories, even the intellectual euphoria about how everything fits together in the end. Baumann (2019) characterizes it as "aesthetically pleasing" that the inevitability of the laws of Nature can be shown using a handful of fundamental principles. At this point, it should be emphasized once again that inevitability must be read in a weak sense. Of course, the laws of nature could be different (perhaps they are in other universes), and the fundamental principles or axioms of our world cannot be proven. However, based on their postulate, clear and inevitable consequences can be derived, such as the nature of the gauge forces from the postulated $SU(3) \times SU(2) \times U(1)$ symmetry of the Standard Model.

We have seen that the criterion of inevitability—when retrospectively applied to constructed theories—is emotionally charged. Nevertheless, we want to regard the associated aesthetics somewhat as an "epiphenomenon" in theory evaluation in fundamental physics: it is caused by the physicist's retrospective amazement at the consistency of the theoretical edifice, ultimately about the comprehensibility and simplicity of the world. However, it does not have a causal effect on theory evaluation—the value is already determined by the inevitable imperative of internal consistency. The aesthetics of inevitability is thus by no means a metaphysical principle that physicists use in addition to the principles of their own field. Rather, it seems to be an emotion that accompanies the use of long-established internal principles, and it does not expand the list of their (aesthetic) evaluation criteria.

 \subsection{Outlook on Naturalness}
Simplicity, symmetry, elegance, and inevitability have served as guiding principles in much of the history of physics and sometimes beyond physics, whose interpretation regularly underwent changes in scientific aesthetic ideals. We have seen that these demands also shaped the rough outline of the BSM physics research programme and that this somewhat subjective vocabulary must indeed be attributed objective significance, even the working method of theoretical physics. However, their significance diminishes compared to the fifth criterion, which appears in particle physics and has therefore been repeatedly discussed separately: naturalness\footnote{See Hossenfelder (2021) for an introduction.}. Naturalness is considered the central concept, "if one has to summarize in one word what drove the efforts in physics beyond the Standard Model of the last several decades" (Giudice 2017, p.\,3).

It is disputed whether the demand for naturalness is physically justifiable. (Dijkstra 2019) summarizes various further positions of the naturalness debate, according to which naturalness is an aesthetic, philosophical, or sociological-heuristic criterion for theory evaluation. The formulations of Dirac's (1974) and Technical Naturalness (Susskind 1979) reminiscent of the idealized simplicity discussed earlier, as both demand a world that is simpler than it appears to us in many places (in the form of unnatural physical constants and the violation of requirements for a universe described only by effective theories). The 't Hooftian formulation (1979) includes a symmetry principle for restoring naturalness. Thus, all variants are at least related to the various desires for simple theories. We refer to the numerous source from the introduction to gain deeper insights into the naturalness problems.

\section{Synthesis: Aesthetics and the Future of Particle Physics}

 \begin{figure}[h!]
  \centering
  \includegraphics[width= 0.80\textwidth]{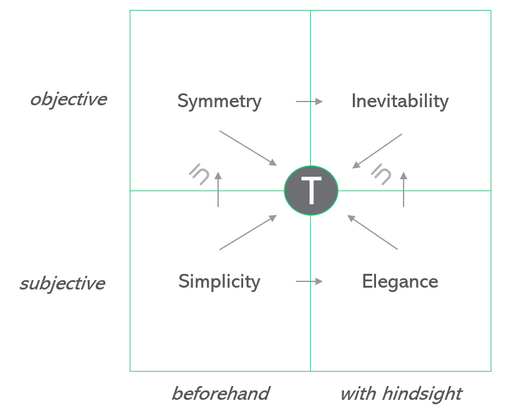} 
    \caption{
The four criteria of beauty have different degrees of objectivity and can be used in the creation and development of a theory $T$ (ex-ante) or in the subsequent theory evaluation after discovering its potential (ex-post). The desire for unification lies at the intersection of simplicity and symmetry as an essential driver of theory formation. Conceptions of simplicity serve as guiding principles for formulating the theory and thus influence the aesthetic content of the theory beforehand. Characterizing for the impression of elegance and inevitability is the element of surprise, which is experience-based and represents a positively disappointed expectation of the physicist. Aesthetic delight arises with hindsight. Theories with symmetries as first principles are a subset of theories considered simple, inevitability arises from a subset of theories considered elegant. It deduces the necessity of certain laws of nature from the mathematical symmetries to which aesthetic value was previously attributed. Both subsets are mathematically precisely definable and thus lie within the objective realm of the space of aesthetic theories. Lastly, symmetric elements in theories are an objective subset of simple elements, theories later understood to be inevitable, given the assumptions, form the objectively aesthetic subset of elegant theories. \label{beauty}}
\end{figure}
 
 The four aesthetic criteria thoroughly examined in Chapter 2 have been found to be closely interconnected rather than isolated from each other, ultimately attributable to the concept of simplicity: Symmetric theories have been identified as a subset of simple theories in which simplicity is objectifiable through a mathematical definition of symmetries. From the ontological parsimony considered in the design of the theory (simplicity), a wealth of phenomena can emerge after the construction and analysis of the theory (elegance). They are two sides of the same coin - the view, both ex-ante and ex-post, of the same construction principle. Elegance and inevitability are the lessons that can be drawn from a theory already aesthetically constructed. Their aesthetic appeal is derived from the same aesthetics previously envisaged in simplicity and symmetry. Thus, these criteria are secondary and always depend on simplicity, by which a theory has already been preferred. The relationships of the four criteria are illustrated in Fig.\;\ref{beauty}.

In particular, the drive for simplicity, stemming directly from the reductionist, unifying approach of fundamental physics, makes this field particularly susceptible to the integration of aesthetic considerations into scientific progress, while most natural science disciplines address the complexity of the world. With the claim of fundamental physics to push the thoughts of the reducibility of the world as far as the human mind allows, there is always a desire for simple first principles. Thus, particle physics is more prone than other disciplines to get lost in dead ends when aesthetic principles mistakenly serve as a guide, especially considering that they all ultimately trace back to simplicity.

While a classic argument for the empirical success of simple theories is that simplicity of phenomena results in a simple theory, this argument is unjustified because simplicity and the criteria lack a property we want to call \textit{metatheoretical permanence.} While the effectiveness of mathematics relies on long-term stable, consistently successful application for adequate prediction of phenomena, the limited efficacy of aesthetic principles is due to the changing nature of aesthetic ideals\footnote{The change in aesthetic ideals is extensively described in McAllister's (1996) book, but it is based on a significantly broader definition of aesthetic criteria, including the aesthetic appeal of intradisciplinary analogies, everyday language metaphors, and metaphysical allegiance.}.
 
 A future metatheoretical permanence of one definition of simplicity is implausible in theory change, but also in the changing metaphysical conceptions of the world—its future conception is unpredictable. The various historical views of simplicity, between Plato and Democritus, between Newton and Leibniz, between their syntactic and ontological nature, suggest that the last word on simplicity is not spoken even today. Thus, aesthetics for theory evaluation will not be subordinate to empiricism for all time. It does not in any way provide a reliable contribution to objective theory evaluation, even if the aesthetic criteria are objectively formulable.

\section{Conclusion}

The disturbing tendency towards the aesthetics of theories, even when confirmed theories are available, is reason enough to incorporate the debate on non-empirical criteria for theory evaluation into the philosophy of science. This is by no means a passing trend but was already revealed, e.g., in 1956 in a confession by Weyl: "My work has always tried to unite the true with the beautiful and when I had to choose one or the other, I usually chose the beautiful"\footnote{In an obituary by Freeman J. Dyson in Nature, March 10, 1956.}. We have analyzed four central concepts that can evoke such an aesthetic sense in the community of foundational physicists: simplicity, symmetry, elegance, and inevitability. Here, a close interconnection of the criteria emerged, as well as a reduction of all aspects to the simplicity of a theory. The assumption that simple phenomena lead to simple theories - regardless of the specific conception of simplicity - makes foundational physics more vulnerable than any other discipline to follow aesthetic guiding principles. The experimental null results of particle physics beyond the Standard Model, which have persisted for several decades, reinforce the urge to turn to new forms of theory evaluation.

While the four criteria may be necessary conditions, they are not sufficient, apart from inevitability: vice versa, adherence to, for example, elegance as in the case of SUSY, can lead to the paralysis of research programs. Simplicity and elegance remain subjective criteria subject to permanent shifts in meaning and therefore lack \textit{metatheoretical permanence}, but they include essential aspects that must be considered in modern physics for a substantive and accurate theoretical description of fundamental phenomena. The idea of simplicity, which is considered here in detail but is not the only one, undergoes objectification in its mathematical formulation via symmetry principles. In advance, fundamental mathematical symmetries are not yet a sufficient criterion to adhere to a theory. It is only in the subsequent consideration as principles from which the encountered natural laws emerge consistently and inevitably that an appropriate modus operandi of foundational physicists lies. The upper-right quadrant of the developed framework is thus the actual desirable criterion. However, on the way there, one also encounters simplicity, elegance, and symmetry - but their adherence at least has the potential to hinder the progress of BSM physics. There may be other reasons why this research programme has made no progress for several decades, including the limited experimental reach of particle accelerators and mathematical inconsistencies of quantum field theory. Nevertheless, cautious handling of aesthetic criteria in theory development and evaluation is a necessary step to reduce the number of possible reasons of standstill.

\subsection*{Acknowledgements}

We thank Ulrich Krohs for valuable discussions and for a critical review. \\ \\ 
\noindent \textbf{Bibliography} \\ \\
\begin{small}
\noindent Bain, J.: \textit{Why Be Natural?}. Foundations of Physics 49(9), 898-914, 2019 \\ 
Banks, T., Seiberg, N.: \textit{Symmetries and Strings in Field Theory and Gravity}, Phys. Rev. D 83, 2011 \\
Boltzmann, L.: \textit{Vorlesungen zu Maxwells Theorie der Elektrizität und des Lichts.} Leipzig : Metzger und Wittig. Bd. II, 1893 \\ 
Boullart, K.: \textit{Mathematical Beauty as a Metaphysical Concept: The Aesthetic of Rationalism. In: G.W. Leibniz-Gesellschaft, ed. Leibniz: Werk und Wirkung}. IV. Internationaler Leibniz-Kongreß: Vorträge. Hannover, G.W.\,Leibniz-Gesellschaft, 69-76, 1983. \\
Branahl, J.: \textit{The Higgs Mass and a Perspective on Broken Autonomy of Scales}. Preprint, 2024\\ 
Carroll, N.: \textit{Philosophy of Art. A Contemporary Introduction}. London: Routledge, 1999 \\
Chakravartty, A.: \textit{Ontological Priority: The Conceptual Basis of Non-Eliminative, Ontic Structural Realism}. In: Landry and Rickles: 187-206, 2012 \\ 
Craig, N.: \textit{Naturalness. A Snowmass White Paper}. Submitted to the Proceedings of the US Community Study on the Future of Particle Physics. arXiv:2205.05708. 2021 \\ 
Derkse, W.: \textit{On Simplicity and Elegance.} Delft: Eburon, 1992 \\ 
Dijkstra, C.: \textit{Naturalness as a reasonable scientific principle in fundamental physics.} 2019 \\ 
Dine, M.: \textit{Naturalness Under Stress Annual Review of Nuclear and Particle Science}. 34 pages.
arXiv:1501.01035, 2015 \\ 
Dirac, P.: \textit{The relation between mathematics and physics}. Proceedings of the Royal Society (Edinburgh) 59, 122-129, 1939 \\
Dirac, P.: \textit{Cosmological Models and the Large Numbers Hypothesis}. Proceedings of the Royal Society of London A. 338(1615), 439-446, 1974 \\
Fischer, E.: \textit{Naturalness and the Forward-Looking Justification of Scientific Principles}, accepted for publication in Philosophy of Science, 2020 \\ 
Franklin, A.: \textit{Whence the Effectiveness of Effective Field Theories?} British
Journal for the Philosophy of Science 71(4), 1235-1259, 2020 \\ 
Gamow, G.: \textit{My World Line}. Viking Press 1970 \\
Gell-Mann, M.: \textit{The Eightfold Way: A Theory of Strong Interaction Symmetry} (Report). California Institute of Technology Synchrotron Laboratory, 1961 \\  
Giudice, G.F.: \textit{Naturally speaking: The naturalness criterion and physics and LHC.} arXiv:0801.2562, 2008 \\ 
Giudice, G.F.: \textit{The Dawn of the Post-Naturalness Era}. Contribution to the volume \textit{From My Vast
Repertoire - The Legacy of Guido Altarelli}. arXiv:1710.07663, 2017 \\ 
Haldane, J.B.S.: \textit{Science and Theology as Art-Forms}, 1927. In J.B.S. Haldane, On Being the Right Size and Other Essays. John Maynard Smith (Ed.). Oxford University Press, 32-44, 1985 \\
Haag, R., Łopuszański, J.T., Sohnius, M: \textit{All possible generators of supersymmetries of the S-matrix}. Nuclear Physics B. 88(2), 257-274, 1975 \\
Hawking, S.: \textit{Gödel and the End of Physics}, public talk at Texas University, March 2002 \\ 
Hossenfelder, S.: \textit{Das hässliche Universum.} S. Fischer Verlag, 2018\\
Hossenfelder, S.: \textit{Screams for Explanation: Finetuning and Naturalness in the Foundations of Physics} Synthese 198, 3727-3745, 2021 \\ 
Ibáñez, L.E. , Uranga, A.M.: \textit{String theory and particle physics. An introduction to String Phenomenology}. Cambridge University Press, 2012 \\ 
Kallosh, R. et al: \textit{Gravity and global symmetries}, Phys. Rev. D 52, 1995 \\
Kragh, H.: \textit{Dirac: A Scientific Biography}. Cambridge University Press, 1990 \\ 
Ladyman, J.: \textit{What Is Structural Realism?} In: Studies in History and Philosophy of Science Part A, 29(3), 409-424, 1998 \\ 
Lakatos, I.: \textit{The Methodology of Scientific Research Programmes.} Edited by John
Worrall and Gregory Currie. Cambridge University Press, 1978. \\
Lisi, G.: \textit{An Exceptionally Simple Theory of Everything}. arXiv:0711.0770, 2007 \\ 
Mach, E.: \textit{The Science of Mechanics: A Critical and Historical Account of Its Development}, 1886. Translation: Thomas J. McCormack, 6. Ed., La Salle, III., 1960 \\ 
Martin, S.P.: \textit{A Supersymmetry Primer}. Perspectives on Supersymmetry. Advanced Series on Directions in High Energy Physics 18, 1-98, 1997 \\ 
Maudlin, T.:  \textit{On the Unification of Physics}, The Journal of Philosophy
93(3), 129-144, 1996 \\
McAllister, J.W.: \textit{Truth and Beauty in Scientific Reason.} Synthese 78(1), 25-51, 1989 \\ 
McAllister, J.: \textit{Beauty and Revolution in Science}, Cornell University Press, 1996 \\
Morrison, M.: \textit{Unifying Scientific Theories: Physical Concepts and Mathematical Structures.} Cambridge University Press, 2000 \\ 
Noether, E.: \textit{Invariante Variationsprobleme}. Nachrichten von der Gesellschaft der Wissenschaften zu Göttingen. Mathematisch-Physikalische Klasse, 235-257, 1918 \\
Penrose, R.: \textit{The role of aesthetics in pure and applied mathematics.} Bulletin of the Institute of Mathematics and its Applications, 10, 266-271, 1974 \\ 
Pollitt, J.J.: \textit{The Ancient View of Greek Art: Criticism, History, and Terminology}. New Haven: Yale University Press, 1974 \\
Quanta Magazine: \textit{Why the Laws of Physics Are Inevitable.}, 2019 \\ 
Quine, W.V.O.: \textit{On Simple Theories of a Complex World}. In: The Ways of Paradox, New York: Random House, 1966 \\
Ramond, P.: \textit{Dual Theory for Free Fermions}. Phys. Rev. D. 3, 2415–2418, 1971 \\
Rhodes, G. et al.: \textit{Facial symmetry and the perception of beauty}.  Psychonomic Bulletin and Review 5, 659-669, 1998 \\
Salimkhani, K.: \textit{Explaining unification in physics internally}, Synthese 198, 5861–5882, 2021 \\
Senjanovic, G.: \textit{Neutrino mass and grand unification.} Lecture at LMU, 2014 \\
 Sober, E.: \textit{What is the Problem of Simplicity?} In: Zellner et al. (eds.), 13-31, 2001\\
Susskind, L.: \textit{Dynamics of spontaneous symmetry breaking in the Weinberg-Salam theory}. Phys. Rev. D
20, 2619, 1970 \\
Taylor, A.M.: \textit{Imagination and the Growth of Science.} London, Murray, 1966 \\
Thagard, p.\,: \textit{The best Explanation}. Journal of Philosophy 75(2), 1978 \\
't Hooft, G.: \textit{Naturalness, Chiral Symmetry, and Spontaneous Symmetry Breaking}. In Proc. of
1979 Cargèse Institute on Recent Developments in Gauge Theories, New York, Plenum Press, 1980\\ 
van Fraassen, Bas: \textit{The Scientific Image}. Oxford University Press, 1980 \\ 
von Neumann, John: \textit{The Mathematician}. In: von Neumann,
John. Collected Works 1. Edited by Abraham H. Taub. Oxford: Pergamon Press, 1-9, 1947 \\ 
von Neumann, John: \textit{Method in the Physical Sciences}. In Leary, Lewis (ed.). The Unity of Knowledge. N.J.: Garden City, 1955 \\
Wallace, D.: \textit{Naturalness and Emergence}. In: The Monist 102.4, 499–524, 2019 \\  
Watson, J.D.: \textit{The Double Helix. A Personal Account of the Discovery of the Structure of DNA.} Critical edition, edited by Gunther S. Stent. London: Weidenfeld and Nicolson, 1968 \\ 
Weinberg, S.: \textit{Dreams of a Final Theory}. Pantheon Books, New York, 1993 \\ 
Wells, D.W.: \textit{The Utility of Naturalness, and how its Application to Quantum Electrodynamics envisages
the standard model and Higgs Boson}. Studies in History and Philosophy of Modern Physics, 49: 102–108, 2013 \\  
Wigner, E.: \textit{The unreasonable effectiveness of mathematics in the natural sciences.} Communications on Pure and Applied Mathematics. 13(1), 1–14, 1960 \\ 
Williams, P.\,: \textit{Naturalness, the autonomy of scales, and the 125 GeV Higgs.} Studies in History and
Philosophy of Science Part B: Studies in History and Philosophy of Modern Physics 51, 82-96, 2015 \\
Williams, P.\,: \textit{Two notions of naturalness}. Foundations of Physics 49(9), 1022-1050, 2019 
 \end{small}
 \end{document}